# Nanostructuring of AlCu alloy NPs obtained by FLASiS[†]

Jesica M. J. Santillán[1], David Muñetón Arboleda[1], Osvaldo Fornaro[2], Marcelo Lester[3], Daniel C. Schinca[1,4], Lucía B. Scaffardi[1]

[1] Centro de Investigaciones Ópticas (CIOp), (CONICET-CIC-UNLP), Camino Centenario y 506, 1897 Gonnet, La Plata, Argentina

[2] Instituto de Física de Materiales Tandil-IFIMAT (UNCPBA), Centro de Investigaciones en Física e Ingeniería del Centro de la Provincia de Buenos Aires-CIFICEN (UNCPBA, CICPBA, CONICET), Pinto 399, B7000GHG Tandil, Argentina

[3] Instituto de Física Arroyo Seco - IFAS (UNCPBA), Centro de Investigaciones en Física e Ingeniería del Centro de la Provincia de Buenos Aires-CIFICEN (UNCPBA, CICPBA, CONICET), Pinto 399, B7000GHG Tandil, Argentina

[4] Facultad de Ingeniería, UNLP, 115 y 49, 1900 La Plata, Buenos Aires, Argentina

Corresponding author: lucias@ciop.unlp.edu.ar and jesicas@ciop.unlp.edu.ar

**Abstract**

In the nanometric scale, alloy nanomaterials behave as new materials since they are able to display structures and properties which are different from those of the pure metal nanostructures or bulk alloys. In this work, the compositional characteristics of AlCu alloy nanoparticles obtained by [†]Femtosecond Laser Ablation Synthesis in Solution (FLASiS) on solid targets with different percentages of Al and Cu were studied. Morphology, structure, composition and plasmonic behaviour of the nanoparticles were analysed by Scanning

Electron Microscopy (SEM), Transmission Electron Microscopy (TEM), Energy-Dispersive X-ray Spectroscopy (EDS) and Optical Extinction Spectroscopy (OES).

From the obtained results, underlying formation mechanism and a configuration model of a host alumina matrix with copper inclusions is proposed. Analysis of the plasmonic resonance spectral region in the extinction spectra of the synthesized colloids using this model show that the nanoparticles have a Cu filling factors of about 0.02 which agrees with those determined independently from TEM images.

The amount of Al and Cu components in the nanoparticle depends on their relative percentages in the initial solid target. Here we describe the possible kinetic formation mechanism of these nanostructures.

## 1. Introduction

Properties of metallic systems span a wide range of applications in different fields of materials science and technology, especially when mixtures of elements to generate intermetallic compounds and alloys are concerned. In some cases, these properties present an enhancement in alloy systems due to synergistic effects, and the rich diversity of compositions and structures, allowing interesting applications in electronics, engineering, biomedicine, catalysis, among others.[1-5] For the case AlCu alloys[6] used mechanical milling for alloying these constituent elements. This technique is appropriate when conventional melting techniques cannot be used due to significantly different metal melting points or immiscibility in the solid state. Inherent properties depend on the amount of each metal within the alloys. For example, Al-rich AlCu alloys yield age-hardening and oxide-dispersion strengthening mechanisms, while Cu-rich AlCu alloys offer very good corrosion resistance, mechanical, thermal, and electrical properties.

In the nanoscale, nanoalloy systems may be considered as new materials since they are able to display structures and properties due to finite size effects, which are distinct from those of the pure metal nanostructures or bulk alloys. There are some metals which are immiscible in the bulk scale but readily mix in finite clusters such as iron and silver.[7]

AlCu nanoalloys have been widely applied in catalysis and machining due to the improvement of material properties.[8] Mechanical alloying,[6,9,10] equal channel angular pressing,[11] high-pressure torsion,[12] sputtering,[13,14] plastic deformation,[15] and laser ablation,[16,17] are some of the synthesis techniques that have been used for producing AlCu alloy nanoparticles (NPs).

Mechanical alloying is the most frequently used technique for the synthesis of AlCu nanostructures, which consists of repeated cold welding, fracturing, and re-welding of blended powder particles in a high-energy ball mill to produce a homogeneous material. Chittineni et al.[6] synthesize AlCu nanostructures by this mechanism and morphologically characterized the grain alloys by Scanning Electron Microscopy (SEM) and X-Ray Diffraction (XRD), founding that the particulate material produced by this synthesis route presents a great dispersion of sizes and shapes. Shanmugasundaram et al.[9] track the structural properties of samples obtained in this way by correlating strength-grain size from a Hall-Petch relation analysis. Here they found that the grain size is a fundamental parameter (although not the only one) that governs the strength characteristics of AlCu nanostructures. On the other hand, Gomez-Villalba et al.[10] perform a more localized analysis of the shape and structure of the grain by Transmission Electron Microscopy (TEM), obtaining information about the assembly mechanism of the nanostructures. In this way, the type of crystallinity that nanometric grains acquire after mechanical alloying can be observed. Similarly, Ammar et al.[18] studied the morphological and strengthening characteristics on larger scales (millimeter range) of samples synthesized by mechanical alloying using Stress-Strain Microprobe (SSM) system and its Automated Ball Indentation (ABI), obtaining results similar to those by Gomez-Villalba, even at larger scales of the analysis.

Another technique used for the synthesis of AlCu nanostructures is the Equal Channel Angular Pressing (ECAP), with which arrays with larger grain sizes are obtained (the smallest are between 80 and 100 nm). Li et al.[11] performed an analysis of the deformation mechanisms by means of compression tests in nanostructures obtained by this technique. They found that by increasing the concentration of Al on the alloyed nanostructures, the Stacking Fault Energy (SFE) decreases, generating a rearrangement of the size of the grains and thus allowing these structures to enter the nanoscale regime with a significant change in strength. Similar results were found by X. H. An et al.[12] in the physical structural changes of strength and ductility in nanostructures obtained by high-pressure torsion. In this same way Y. Zhang et al.[15] study the effect of SFE, strain rate and temperature on microstructure characteristics of AlCu nanostructures obtained by plastic deformation.

Draissia et al.[13] obtain AlCu films by sputtering from alloyed targets. They found that the strengthening of the film is strongly dependent on the inclusion of Cu. McKeown et al.[14] characterize the nucleation of the AlCu alloy in this type of film in real time using transmission electron microscopy, shedding light on the crystallinity state of the films. On the other hand, Mojumder et al.[19] analysed the physical characteristics of multidimensional nanostructured arrangements of AlCu through molecular dynamics simulations. They found that the strengthening of the structures (Hall-Petch effect) is largely related to the size of the grain and the temperature, setting the degree of array ordering.

Synthesis of different metal alloy nanostructures using nanosecond pulse laser ablation has been postulated as a promising technique, due to the high purity of the samples obtained and the versatility of its implementation.[20,21] Particularly for AlCu alloy, Perez et al.[16] studied the mechanism of nanosecond pulse laser ablation in Al, Cu and the AlCu alloy itself within the initial framework of pulsed laser deposition of thin films. Velocity distributions of the ablated Al and Cu atoms were measured by monitoring their time of flight from the target to the probe laser beam. However, they do not analyse in detail the structural properties of the particulate material produced during ablation. Meng et al.[17] makes a more in-depth computational analysis of the behaviour of the shock wave on the AlCu alloy target during ablation. However, in the ultrashort pulse regime, only laser ablation of pure metals or laser powder bed fusion of AlSi alloys were studied.[22,23] In this framework this paper deals with the generation and characterization of AlCu alloy NPs using Femtosecond Laser Ablation Synthesis in Solution (FLASiS), from substrates that were manufactured by conventional melting and pouring method of the adequate quantity of the alloy elements pure Al and Cu.[24] Their physical and chemical properties rely on a good characterization of the grain size of the alloy, as well as its morphological and structural properties. In the present work we carry out the analysis of the accommodation (structuring) of each alloyed metals within NPs obtained after FLASiS. This internal structure is analysed by means of SEM, TEM and Energy-Dispersive X-ray Spectroscopy (EDS) analysis. Optical Extinction Spectroscopy (OES) is used to obtain the experimental extinction spectra for different alloy composition suspensions. A theoretical analysis of the experimental spectra based on Mie scattering is carried out to understand the dynamics in the formation of the NPs. The theoretical Levenberg-Marquardt (L-M) method[25] is implemented to fit the experimental curves, taking into account a NP dielectric function given by Maxwell Garnet (MG) model of the different percentages of the AlCu alloy.

## 2. Materials and Methods

### 2.1 Target Sample Preparation

Substrates of various chemical compositions were prepared based on the reduced equilibrium phase diagram for AlCu alloy shown in Figure 1. The phase diagram was obtained by using OpenCalphad (OC) software[26] and COST507 free alloy database.[27,28]

Samples were manufactured by conventional melting and pouring method of the adequate quantity of the alloy elements, 99% purity Al and electrolytic Cu.[24] The melting was achieved by using an induction furnace under Ar atmosphere. The induction method lets a correct mixture of the melted liquid before performing the casting. Pouring was realized into graphite molds.

In Figure 1 $\alpha$ phase corresponds to fcc AlCu while $\theta$ phase corresponds to $Al_2Cu$. The amount of Cu used in each sample is indicated by dashed-dotted lines. For Cu weight percentage in the range 0.5 to 4.5 (green lines), the samples are in $\alpha$ phase, while for weight percentage between 12 to 40 (cyan lines), the samples are in compositional mixtures of $\alpha + \theta$ phases and sample of 54 Cu weight percentage (red line) is in intermetallic $\theta$ phase. Al pure (magenta line) and Cu pure (not shown in the Figure) phases are also analysed.

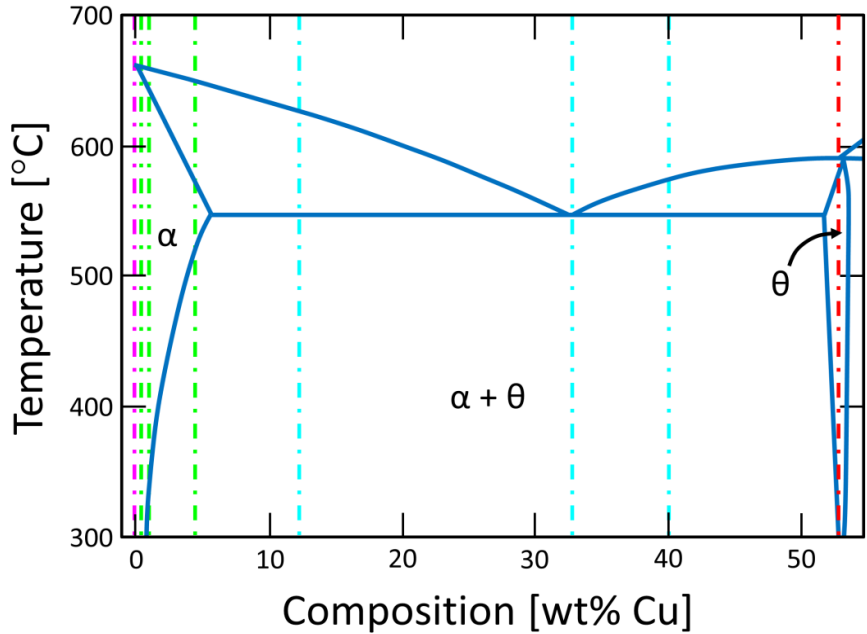

Figure 1: Equilibrium phase diagram of AlCu system alloy. Dashed dotted lines mark the Cu weight % used in the alloy samples.

Table 1 summarizes the prepared samples (1$^{st}$ column), their composition as weight % Cu and atomic % Cu (2$^{nd}$ column) and their corresponding phase in accordance to Figure 1. The alloy composition is expressed as AlCu$_{x\%}$ being x% the Cu weight percentage.

Traditional metallographic method was used to observe the microstructure of the different substrates used in laser ablation process. Surfaces were prepared by mechanical polishing, using water refrigerated SiC paper, followed by diamond and alcohol suspension in a smooth cloth. A final electrolytic polish and etching process is applied to observe the microstructure under an optical microscope. For this purpose, a Leica Metallographic Inverted Microscope DMI-4000 was used.

Table 1: Substrate chemical composition

| CODE (AlCu$_{x\%}$) | Composition | | Detail |
|---|---|---|---|
| | wt% Cu | at% Cu | |
| Al | 0.0 | 0.0 | Pure Al |
| AlCu$_{0.5\%}$ | 0.5 | 0.21 | Homogeneous α T6 solubilized |
| AlCu$_{1.5\%}$ | 1.5 | 0.64 | |
| AlCu$_{4.5\%}$ | 4.5 | 1.96 | |
| AlCu$_{12\%}$ | 12.0 | 5.47 | α + (α + θ)$_{Eut}$ |
| AlCu$_{33.3\%}$ | 33.3 | 17.49 | (α + θ)$_{Eut}$ |
| AlCu$_{40\%}$ | 40 | 22.06 | θ + (α + θ)$_{Eut}$ |
| AlCu$_{53.7\%}$ | 53.7 | 33.3 | Θ |
| Cu | 100 | 100 | Pure Cu |

Characterization and composition of bulk alloys were performed with SEM microscope FEI QUANTA 200 by EDS technique.

**2.2 Colloid preparation FLASiS and characterization techniques**

All alloy samples as well as pure metals were used as solid targets for 120 fs pulse laser ablation 600 µJ in water HPLC. Colloids were obtained by focusing this laser on a 10 mm diameter and 1 mm thick AlCu alloy solid disk target immersed in water. To enable obtaining concentrated suspensions, a 1 cm height water column was left over the target within the ablation vessel.

Morphology, structure, composition and size of the NPs were analysed using a FEI-Talos F200X G2 FEG Scanning Transmission Electron Microscope in STEM mode combined with a High Angle Annular Dark Field (HAADF) detector, with resolution of 0.16 nm. The detector allows registering the transmitted beam electrons that have been scattered by the sample through a relatively high angle. EDS analytical technique was used for the elemental analysis on our samples.

For TEM measurements, colloid samples were prepared few hours before the experiment by drying a drop of the colloidal dispersion on ultrathin carbon film supported on Ted Pella holey carbon copper grid.

A Shimadzu UV-1650PC spectrophotometer (200 nm to 1100 nm wavelength range) was used to obtain absorbance spectra from the as-prepared colloids for different concentrations of AlCu alloy as well as pure Al and pure Cu.

## 3. Results and discussions

Pure Al samples solidify as different grains with a unique composition. For $\alpha$ phase samples (0.5 to 4.5 wt% Cu) the solid could show a micro-segregation pattern, but a T6 standard solution treatment was used to achieve a homogeneous chemical composition.[29-31] The limit of wt% Cu for this phase to be homogeneous (close to 5%) depends on the alloying temperature (Figure 1). For a higher Cu composition samples, it is impossible to avoid the formation of the secondary $Al_2Cu$ $\theta$ phase. For this reason, samples with 12 to 40 wt% Cu are found in $\alpha$ (AlCu) and $\theta$ ($Al_2Cu$) phases mixture. Finally, for samples above 54 wt% Cu correspond solely to θ phase. It is clear then that, for samples with different wt% Cu the latter is not homogeneously distributed since it effectively composes different solid phases.

Alloy microstructures show dramatic changes as wt% Cu increases. Figure 2 is a collection of Optical Microscopy (OM) and SEM images of specific targets with increasing values of Cu content.

Copper distribution corresponds to the expected microsegregation pattern after casting solidification. Figure 2 (a) is an OM image of AlCu$_{0.5\%}$ as-cast structure, without T6 treatment. Panel (b) is a SEM image of AlCu$_{12\%}$ sample, showing the appearance of the θ phase (white colour in the micrograph) and α phase (grey colour in the micrograph). Panel (c) shows an OM image of the AlCu$_{33.3\%}$ sample (eutectic composition). Panel (d) is a SEM image of AlCu$_{40\%}$ sample showing an island-pattern microstructure. Panel (e) is a SEM micrograph of an AlCu$_{53.7\%}$ sample, showing the formation of intermetallic phases with rich-Cu dendrites patterns (light grey) due to decreasing chemical composition of the remaining liquid during the solidification. As can be seen, the structure of a homogeneous solid of θ composition cannot be achieved, due to the fact that temperature formation for intermetallic phase is less than liquid temperature, giving rise to typical to Cu dendrite-like primary phase.

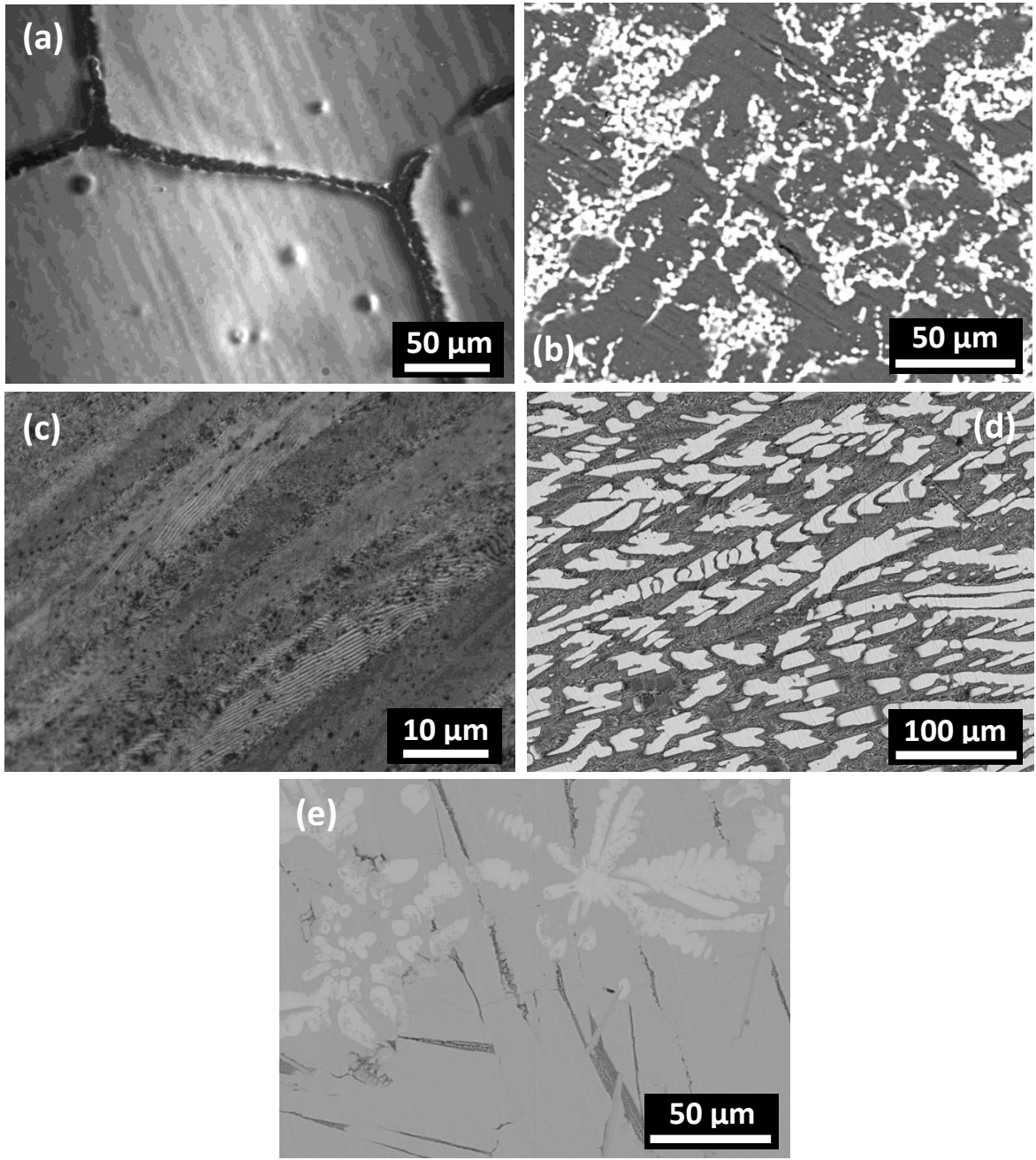

Figure 2: Optical and SEM microcopy images of as-cast microstructures of substrate targets. (a) OM image of AlCu$_{0.5\%}$, without T6 treatment; (b) SEM image of AlCu$_{12\%}$. (c) OM image of AlCu$_{33.3\%}$ (eutectic composition); (d) SEM image of AlCu$_{40\%}$ sample; (e) SEM image of AlCu$_{53.7\%}$.

As can be seen from the sequence, the microstructure was changing with the content of Cu. For Cu-content smaller than maximum solubility limit, a homogeneous solid solution could be obtained through a classic T6-solution heat treatment at 550 °C during 4 hs.

However, above that limit, it is impossible to avoid separating into two distinct and well-identified phases, as fcc (AlCu) α and the intermetallic θ. The quantity of each phase depends on the sequence of solidification.

To identify different present phases in the samples, EDS analysis was performed on specific sites to characterize their microstructures. Figure 3 corresponds to an AlCu$_{12\%}$ alloy. The analysis over the lighter zone, marked by red square, reveals Cu and Al content (46% and 54% respectively, Panel (b)), indicating that is close to the eutectic composition expected from the microsegregation pattern of dendrite growth.

Panel (c) shows the EDS spectrum from the grey zone marked by a green circle. It can be observed a strong peak at the Al energy demonstrating a majority presence of the host metal (96%). The small Cu peak indicates a very low Cu content (4%) in the selected area expected from the α phase.

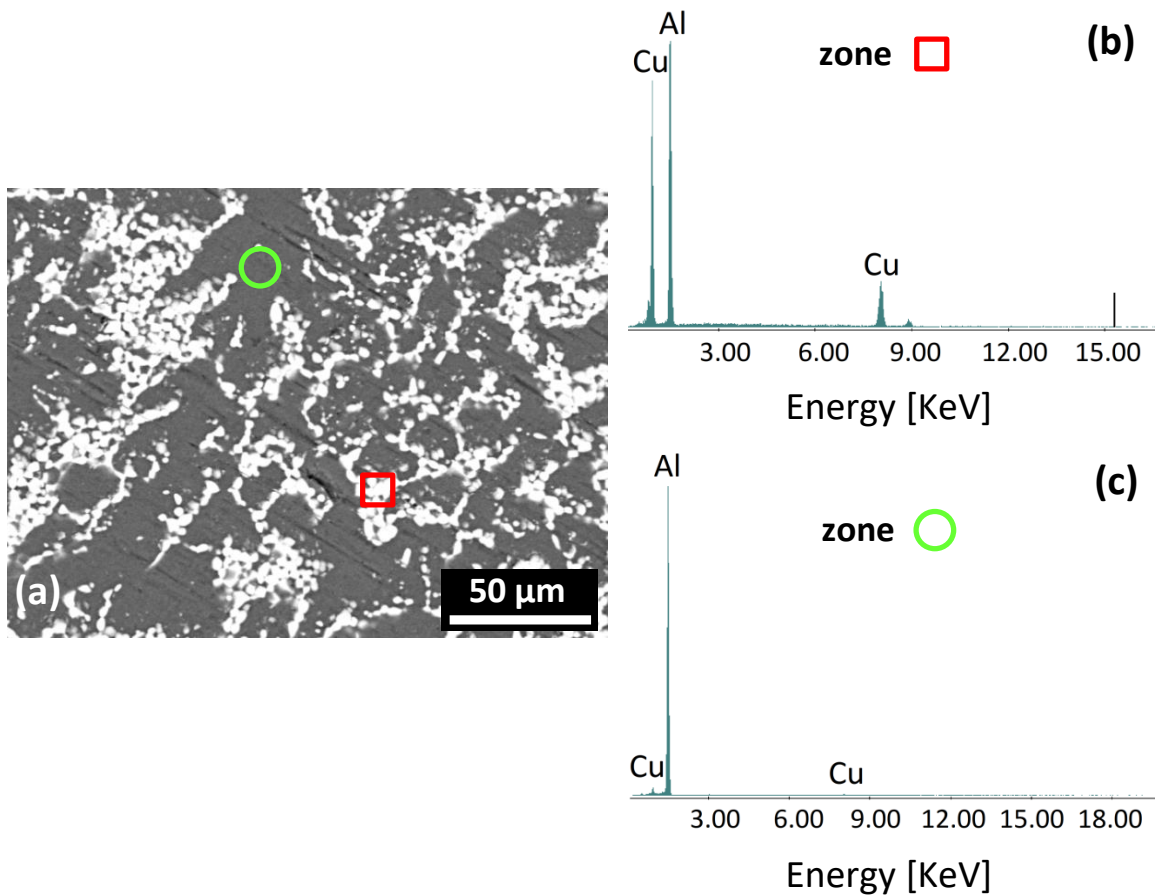

Figure 3: (a) SEM of AlCu$_{12\%}$, (b) EDS corresponding to the red square zone and (c) EDS at the green circle zone in the micrograph. In Panels (b) and (c) X axis shows energy in KeV and Y axis number of counts.

A similar analysis was performed on the hyper-eutectic alloy of AlCu$_{40\%}$ (Figure 4). In this case, the primary phase in solidify is the θ phase, surrounded by eutectic liquid. The EDS analysis was performed on specific sites identified in the micrograph.

The analysis over the lighter zone, marked by pink cross, reveals Cu and Al content in a very approximate ratio of 50% each (Panel (b)). On the other hand, Panel (c) shows the EDS spectrum of the grey zone marked by a green circle. It can be observed that, as opposed to the case shown in Figure 3, in this sample the grey zone has small white spots, suggesting a larger presence of the θ phase as the Cu content increases during sample preparation. The Al and Cu content in this zone is 69% and 31% respectively.

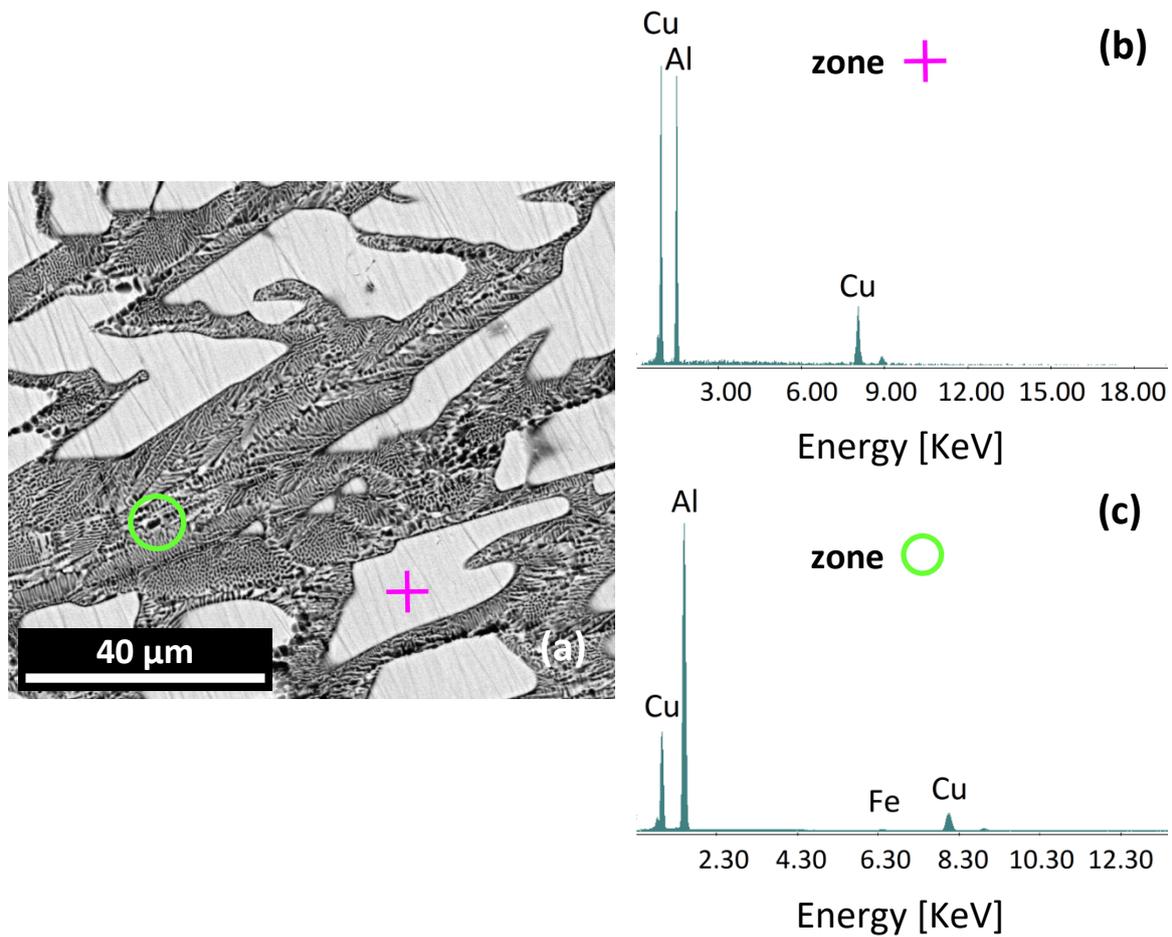

Figure 4: SEM of AlCu$_{40\%}$, (b) EDS corresponding to the pink cross zone and (c) EDS at the green circle zone in the micrograph. In Panels (b) and (c) X axis shows energy in KeV and Y axis number of counts.

Finally, as shown in Figure 5 for the AlCu$_{53.7\%}$, rich-Cu dendrites can form due to the high liquid temperature during preparation. The majority phase is the θ phase. Also, a poor-Cu of approximate α composition could appear in less quantity. Panel (a) shows a SEM micrograph of the sample, exhibiting three distinct zones. Panel (b), (c) and (d) show the EDS analyses for these zones. A plain grey region, marked with green circle, presents a similar content of Al and Cu about 50% each (Panel (b)). The dendritic zone (light grey, marked with a red square) has a higher Cu content (62%) with respect to Al content (38%) as shown in Panel (c). Lastly, the dark grey zones marked with cyan cross (Panel (d)), has a high content of Al host (70%) and 30% of Cu.

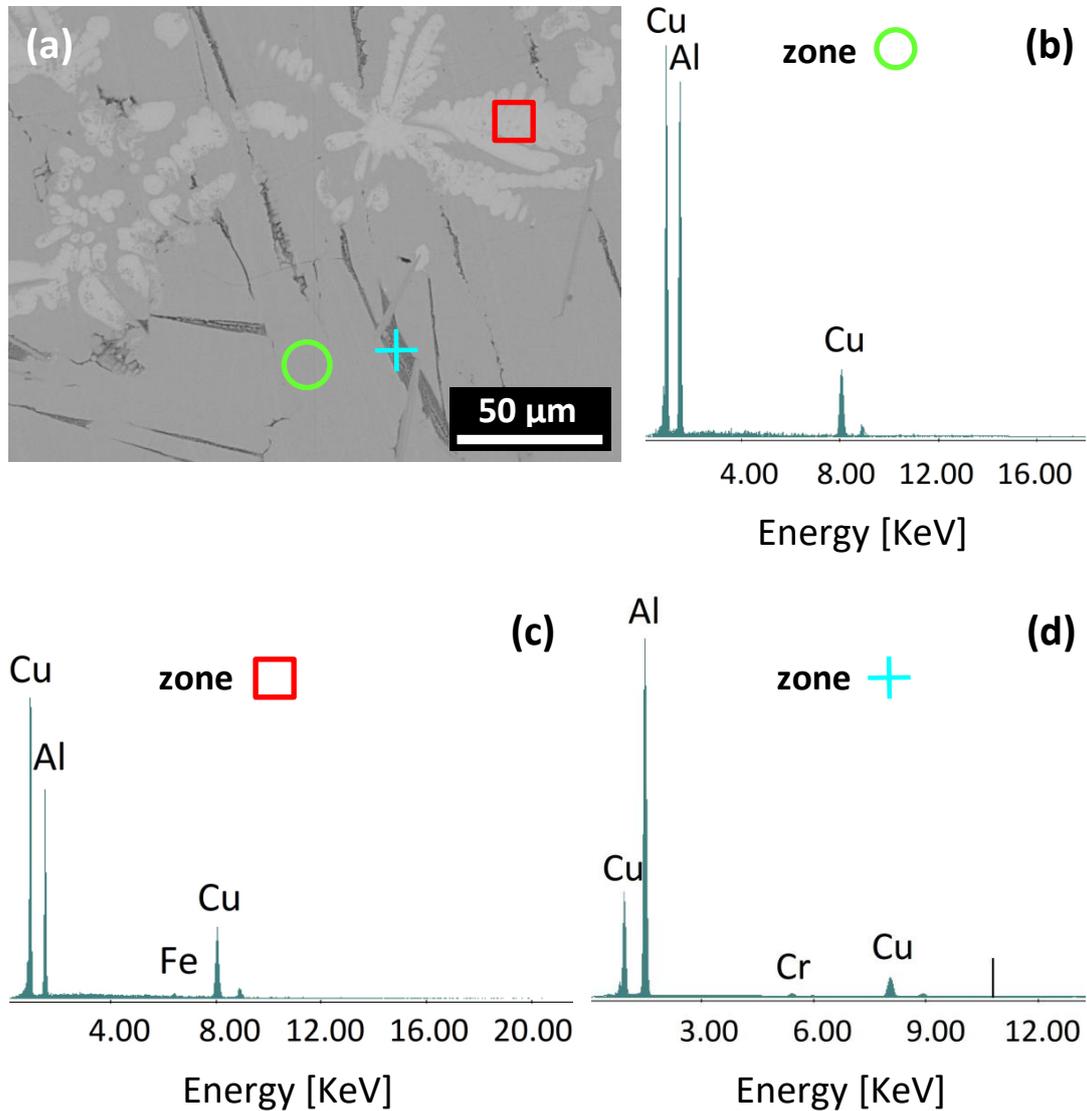

Figure 5: (a) SEM micrograph of AlCu$_{53.7\%}$, (b) EDS corresponding to the green circle zone, (c) EDS at red square zone and (d) EDS at the cyan cross zone in the micrograph. In Panels (b), (c) and (d) X axis shows energy in KeV and Y axis number of counts.

Elemental composition results in Figures 3, 4 and 5 are in agreement with those expected from samples preparation method, shown in the AlCu alloy phase diagram in Figure 1.

To obtain the AlCu NPs colloids all these samples were ablated as described in the experimental section. Figure 6 shows typical results of this process. Panel (a) shows a panoramic view of the target region of an ablation AlCu$_{12\%}$ sample, where the path of the ablation laser beam is observed as parallel straight grooves (meander lines). It can be seen that the laser spot is larger than the inherent size of the microstructure, so ablation affects zones results with different elemental composition. Panels (b) and (c) are enlargements within the grooves for AlCu$_{33.3\%}$ and AlCu$_{40\%}$.

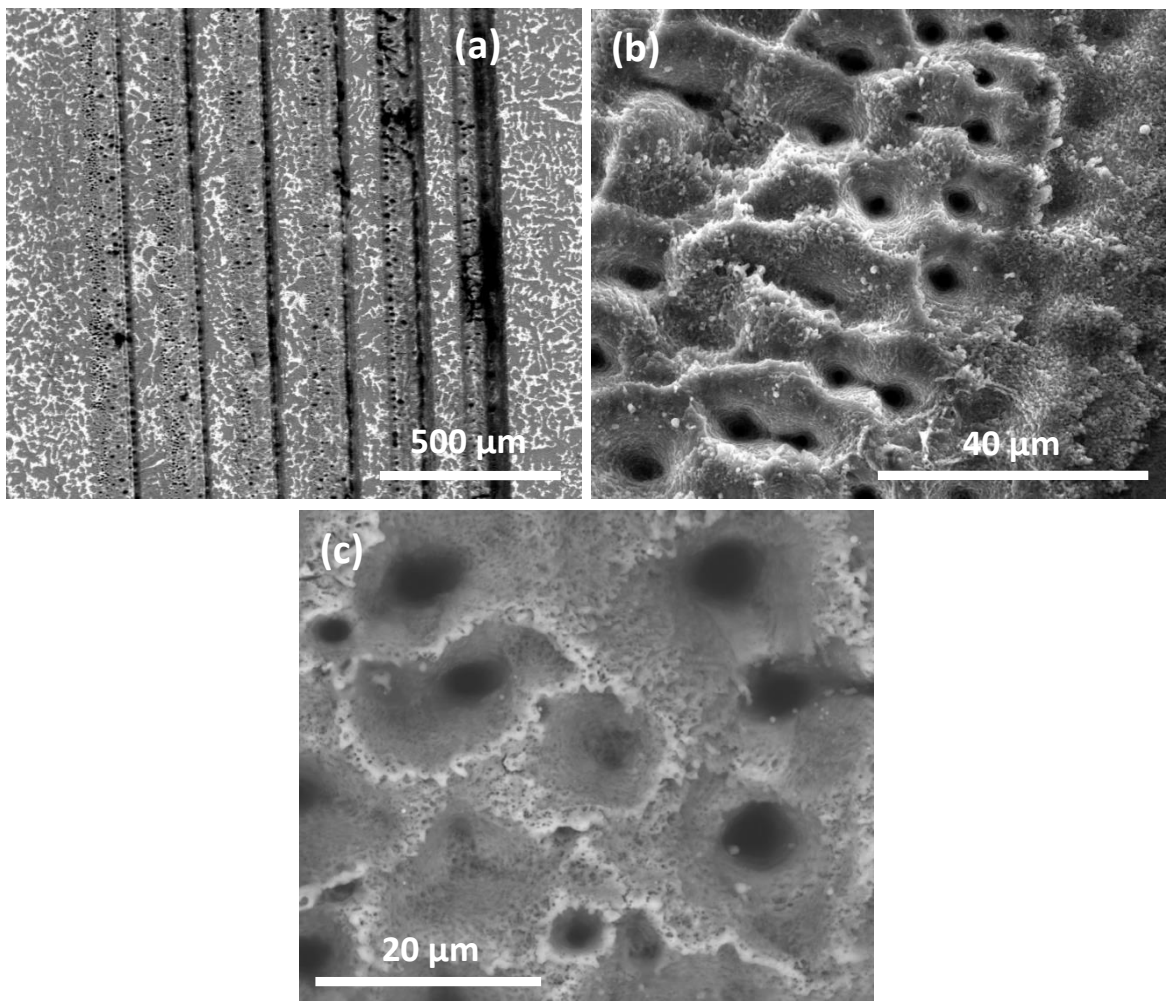

Figure 6: SEM image of ablated surface of (a) AlCu$_{12\%}$ and (b) AlCu$_{33.3\%}$; (c) AlCu$_{40\%}$. Panels (b) and (c) are enlargements within the grooves for AlCu$_{33.3\%}$ and AlCu$_{40\%}$.

Different HADDF images were obtained for the colloids synthesized by FLASiS with different percentages of AlCu alloy samples.

Typical image results for AlCu$_{33.3\%}$ are shown in Figure 7. Panel (a) is a panoramic view showing the spherical morphology of the obtained NPs, spanning a wide range of sizes. It can be seen that there are few NPs in the 200 nm radius range, together with a large number of NPs in the 50 nm radius range and a vast majority of smaller (< 10 nm) NPs. This is also observed in panel (b) with a 100 nm resolution, where it can be seen that the NPs have internal structures (dashed line yellow circles).

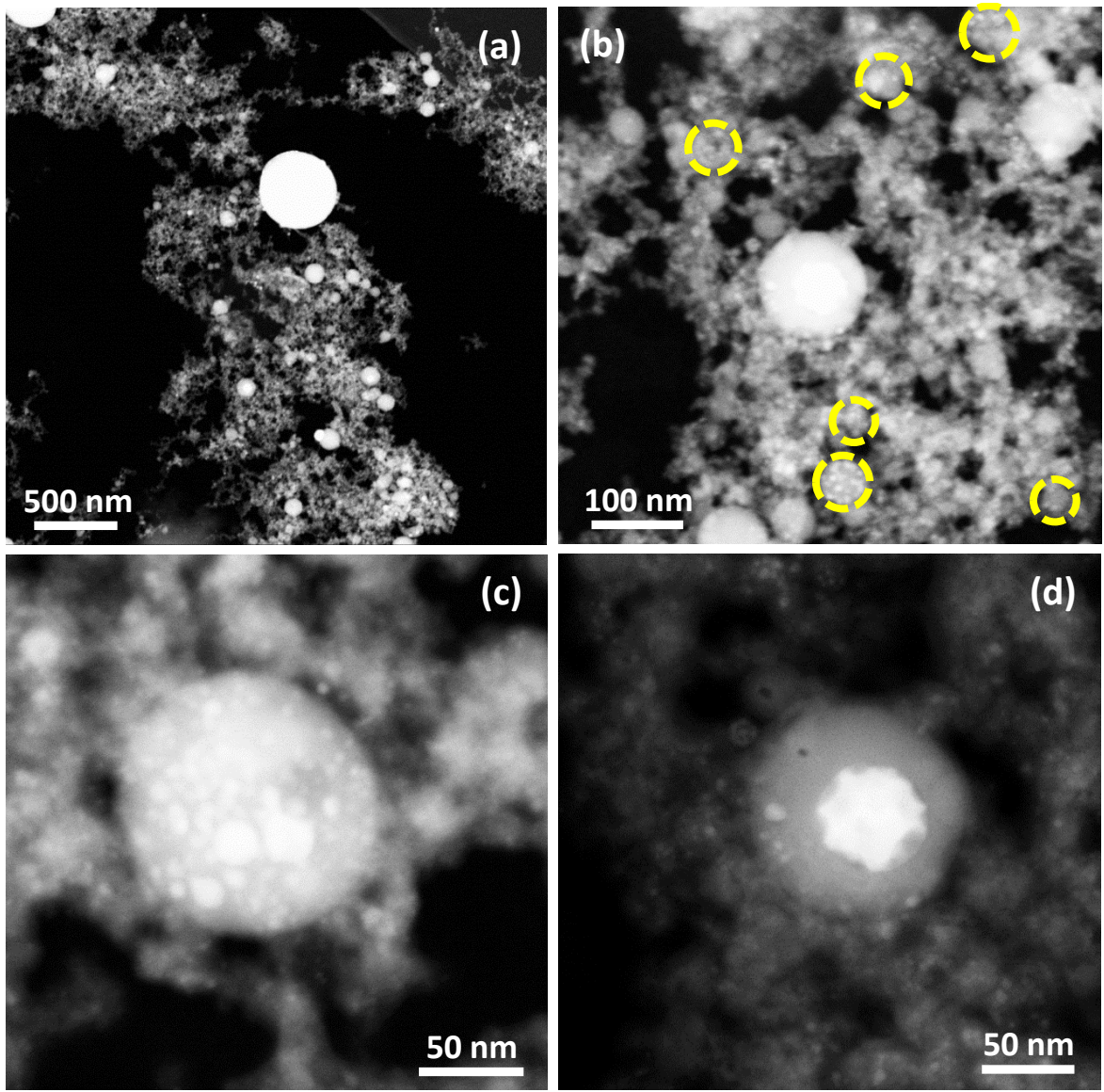

Figure 7: HADDF images of colloid obtained by FLASiS of AlCu$_{33.3\%}$ alloy: (a) Panoramic view of the large NPs; (b) image showing internal structures (dashed line yellow circles) with a 100 nm resolution; (c) single NPs showing details of internal structure; (d) another single NP surrounded by nanopetals-like forms.

A particularly large NP was selected in Panel (c) to shows in more detail its internal structure, whose composition will be analysed later. This type of structure is common also in the smaller NPs. Panel (d) shows another type of structured NP, surrounded by a less dense region in the form of nanopetals. This result is similar to that described by Svarovskaya et al.,[32] during electric explosion of bimetallic Al/Cu wire. They state that nanopetals are similar to oxidation products of electroexplosive Al and Al/AlN nanopowders, all under the same conditions. Notice that in all Panels of Figure 7 an amorphous phase appears as whitish zones surrounding the NPs.

Similar studies were performed on colloidal NPs obtained by FLASiS of AlCu$_{53.7\%}$ bulk alloy. When this sample is ablated, again similar substructured NPs are obtained, as observed in Figure 8 (a). This kind of substructured NPs is also observed in smaller NPs. To find out their qualitative composition and elemental distribution, an EDS analysis was performed on the above selected large NP. Aluminium seems to be uniformly distributed (green dots) over the volume of the NP of about 90 nm radius (Panel (b)) while copper appears to be preferentially located as inclusions (red dots) with size range between 4 nm and 15 nm radius approximately (Panel (c)). Panel (d) shows Al and Cu distributions superimposed within the NP itself. It can be seen the good spatial correspondence between the Cu distribution and the inclusions.

Using a similar fabrication method, Noor et al.[33] studied the thermal-chemical characteristics of AlCu alloy NPs, their reactivity and stability at various temperatures compared with that of pure Al particles of similar dimensions. They found that Al, Cu and O were uniformly distributed. The presence of O was due to the formation of a passivation layer in the form of oxide.

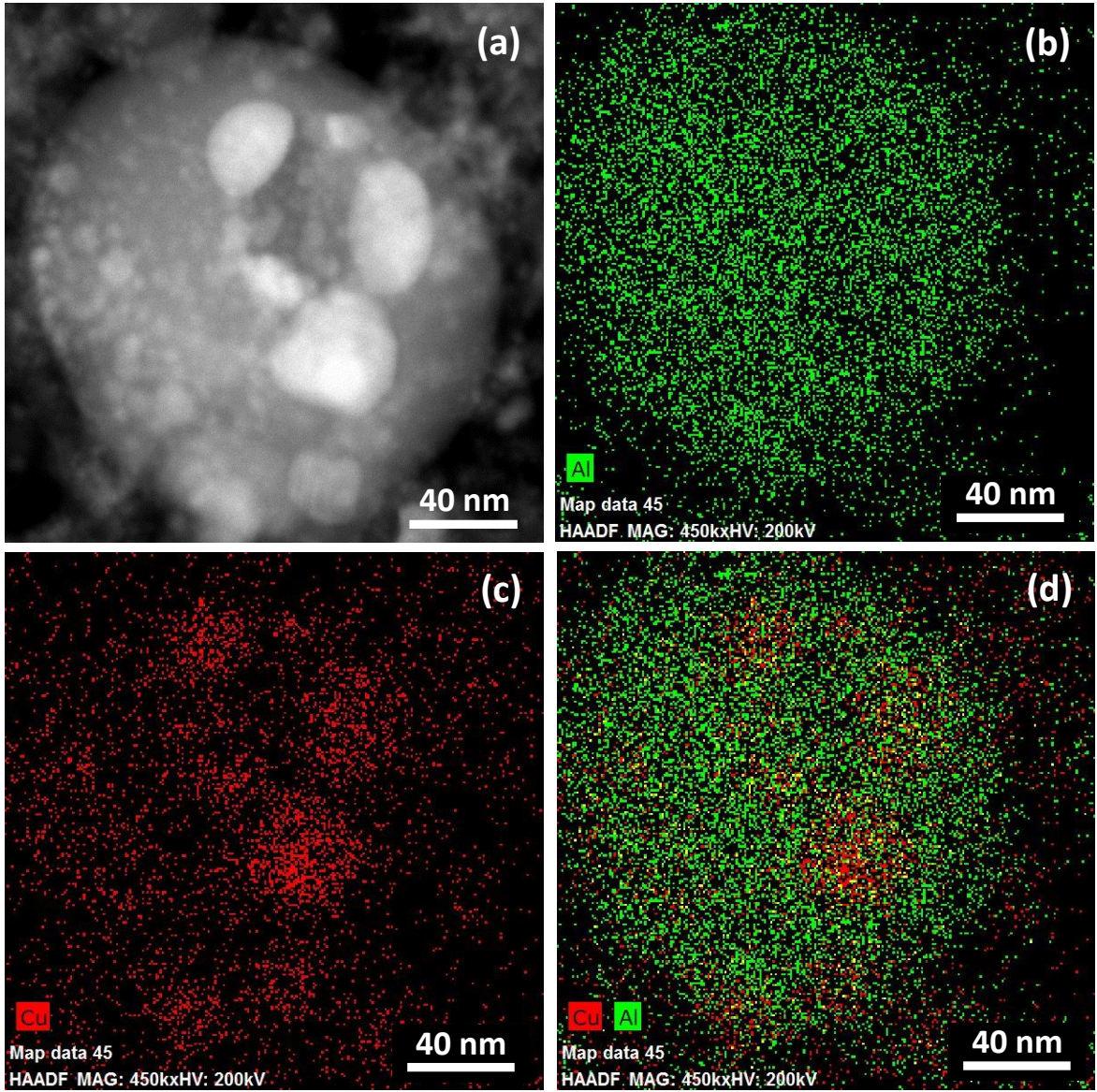

Figure 8: Analysis of colloidal NPs obtained by FLASiS of AlCu$_{53.7\%}$ bulk alloy: (a) structured NP; (b) mapping of Al content distribution in the structured NP; (c) mapping of Cu content distribution in the structured NP; (d) superimposed of Al and Cu mapping.

Recalling that in typical process of NPs formation by FLASiS, nucleation proceeds from the gaseous phase, a gas containing mixtures of Cu, Al, O and H must be expected because ablation proceeds in pure water. The effective chemical composition of this gas strongly depends on the specific target site where ablation takes place in the sample. The plasma composition does not affect the nucleation rate, but the chemical composition of the NPs through a series of reactions. Alumina ($Al_2O_3$) is one of the most stable oxides that may form starting with these atomic species. It may appear from vaporized Al and $O_2$ during the high transient plasma temperatures about 6500 K, forming alumina in one of its allotropic forms. Instead, Cu appears as metallic Cu within of alumina or could combine with free $O_2$ to form $Cu_2O$[34] in water.

When EDS analysis is performed on different sites of the same NP in Figure 8 (a), a quantitative composition can be obtained. Figure 9 (a) shows an $AlCu_{53.7\%}$ alloy NP with non-uniform inclusions. The green circle corresponds to the region analysed by EDS whose spectrum in shown in Figure 9 (b). The presence of C peak corresponds to the carbon atom in the grid.

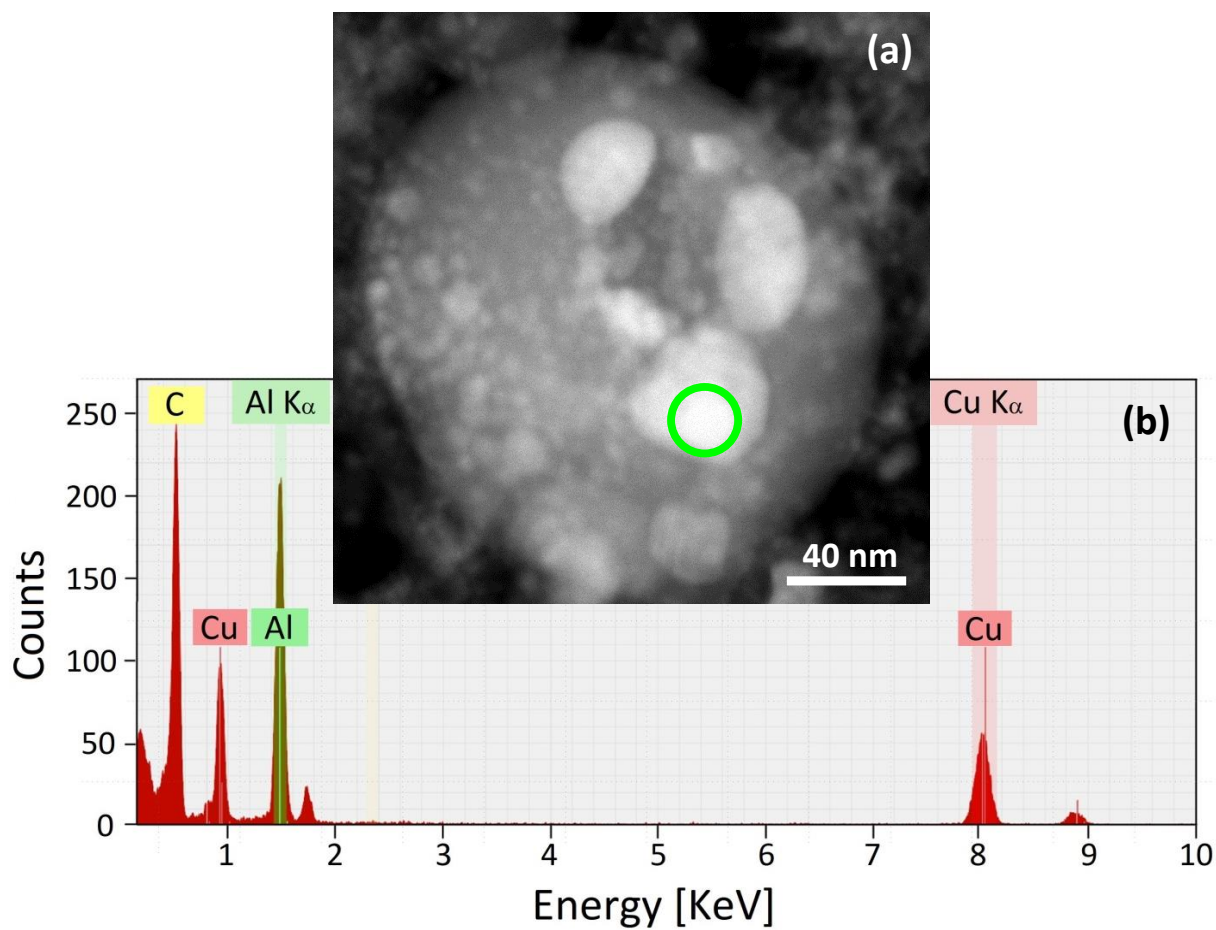

Figure 9: (a) TEM image of a single AlCu$_{53.7\%}$ alloy NP with non-uniform inclusions. The green circle corresponds to the region analysed by EDS; (b) EDS spectrum of the region marked in (a).

Table 2: Elemental composition from the EDS analysis of the NP region shown in Figure 9

| Element | Series | Norm. wt% | Error in wt% |
| --- | --- | --- | --- |
| Copper | K-series | 65.98 | 6.30 |
| Aluminium | K-series | 16.77 | 1.73 |
| Oxygen | K-series | 17.25 | 1.80 |

Table 2 summarizes the quantitative EDS spectrum shown in Figure 9 (b). Examination of element composition in substructural characteristics of inclusions allowed establishing that electron-dense inclusions are copper enriched, while the similar wt% of aluminium and oxygen suggests that Al has undergone an oxidation process. During pulsed laser ablation process, there is a local increase in temperature at the beam focus. Aluminium reacts with water at this temperature producing a porous texture (see Figure 6) within which Cu rich inclusions.

It is important to highlight that after FLASiS the obtained NPs from bulk alloy AlCu$_{53.7\%}$ do not keep the original elemental composition, being the Al 17 wt% and Cu 66 wt% taking into account the presence of oxygen with a 17 wt% approximately (Table 2). However, if only the metals need to be considered, their relative percentage changes to Al 21 wt% and Cu 79 wt% within the NP. To the best of our knowledge this is the first time this result is reported by FLASiS of AlCu NPs.

To complete the elemental composition analysis of this alloy NP, Figure 10 shows EDS results taken from a region away from the inclusions. Panel (a) shows this region marked with a green circle and Panel (b), the corresponding EDS spectrum where a high Al peak is seen.

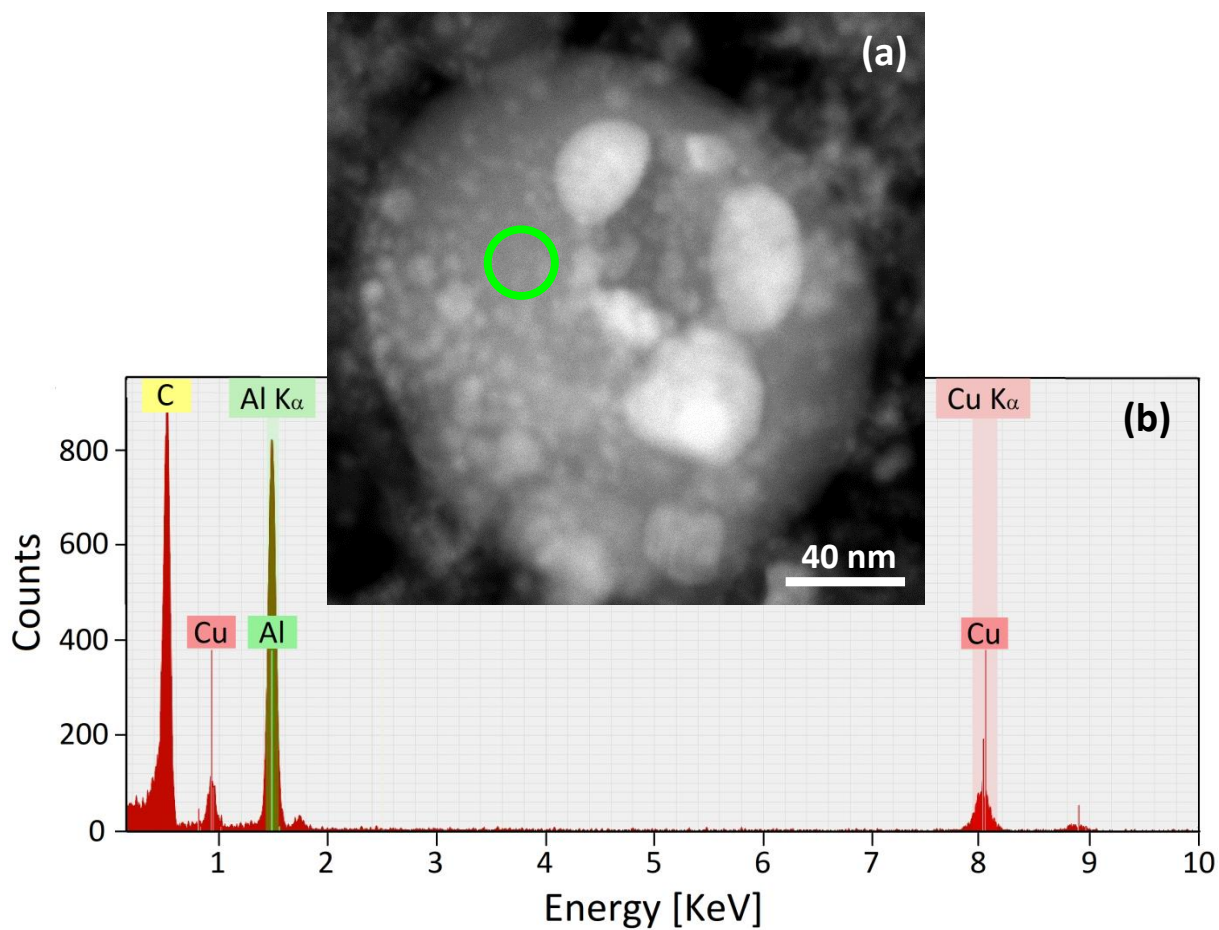

Figure 10: (a) TEM images of a selected alloy NP; (b) EDS spectrum corresponding to the area marked with a green circle in the NP.

Table 3: Elemental composition from the EDS analysis of the NP region shown in Figure 10

| Element | Series | Norm. wt% | Error in wt% |
| --- | --- | --- | --- |
| Copper | K-series | 14.26 | 1.86 |
| Aluminium | K-series | 42.42 | 4.20 |
| Oxygen | K-series | 43.32 | 4.34 |

Table 3 summarizes these quantitative results. Again, the wt% of Al and O are very similar, suggesting once more the presence of these elements as alumina phase in the NP. Besides, the presence of alumina is larger than Cu.

Until now, TEM and EDS analysis allowed us to describe morphology and elemental composition of NPs obtained from AlCu$_{33\%}$ and AlCu$_{53.7\%}$ alloys shown in Figure 7 and Figure 8, respectively. In these Figures it can be observed alumina NPs with Cu inclusions for which filling factor $f$ is determined as the ratio of the volume of the inclusions to the whole volume of the alumina NP. For the selected AlCu$_{33\%}$ NP in Figure 7 (c), $f = 0.015$. Similarly, for AlCu$_{53.7\%}$ NPs, shown in Figure 8 (a), $f = 0.053$. From these results it can be seen that there is a direct correspondence between the filling factor in the NPs and Cu percentage in the bulk samples. Notice that, these $f$ values may not be representative of the total colloidal suspension since there is a vast majority of smaller NPs which may have smaller filling factors and may influence the modelling of the experimental optical extinction spectra.

A complementary study using UV-visible spectroscopy was carried out. Optical extinction measurements of NPs colloids synthesized by FLASiS were recorded for the different AlCu alloy samples mentioned in Table 1, including also pure synthesized Al and Cu for comparison. Figure 11 shows these spectra, where the typical Cu plasmon resonance close to 600 nm is clearly seen. Al NPs are known to have plasmon absorption band close to 190 nm which is out of the spectrophotometer range. As expected, Cu resonance gets smaller as its percentage decreases within the alloy. Under 33.3 wt% Cu, the plasmonic resonance peak is undetectable. Notice that the plasmon peak for pure Cu in water appears around 650 nm, which is typical for Cu-Cu$_2$O core shell NPs,[34] while for alloy NPs appears blue shifted to 600 nm, typical for core Cu without shell.

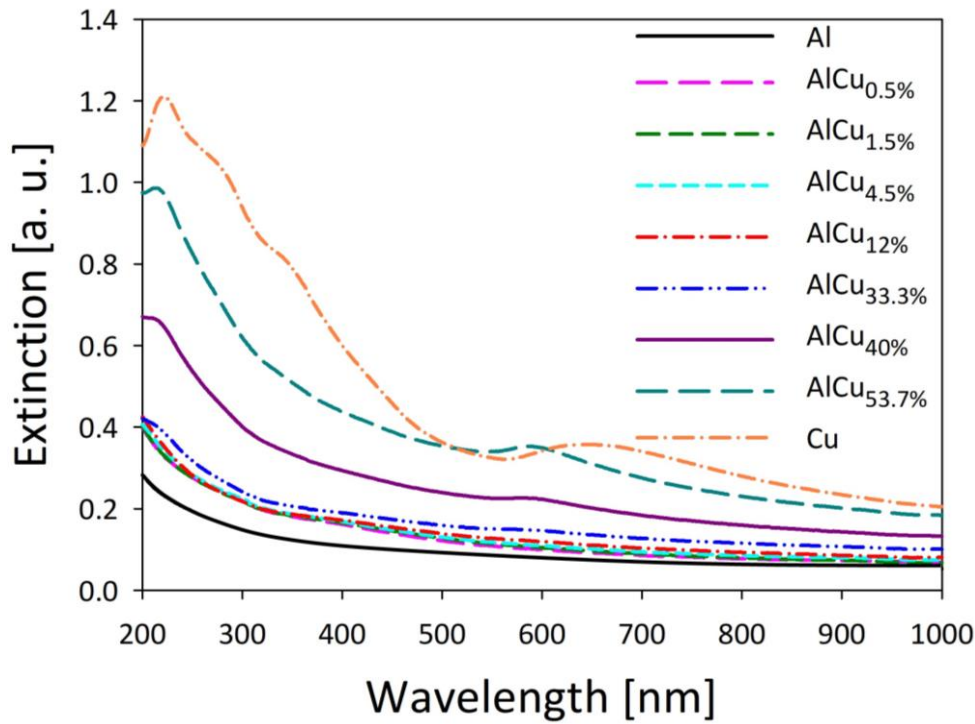

Figure 11: Colloid AlCu alloys extinction spectra for different Cu percentages. Pure Al and pure Cu synthesized by FLASiS in water are added for completeness.

Based on the elemental mapping of Figure 8, a model for the morphological composition of AlCu alloy NPs can be proposed to qualitatively describe the main features of the experimental spectra shown above. As shown in Figure 2, the distribution of Cu and Al in the bulk alloy is not homogeneous. The samples' target surface present different patterns, depending on the Cu concentration in the Al host. Figure 6 shows that the formation pattern of the Al and Cu phases in the prepared alloys, exhibits similar sizes as the laser spot (Panel (a)). So, during laser ablation process, the laser beam impacts over these distinct phases, sputtering out Cu or Al into the surrounding aqueous medium. To delve into the alloy NP model, we need to understand how oxygen in water influences the formation of NPs.

The presence of free oxygen produced during FLASiS in water, together with the EDS results shown in Table 2 and 3 concerning the presence of aluminium and oxygen in very similar weight percentages and the high aluminium affinity for oxygen molecules to form strong bonds with them quickly,[35,36] supports the idea of a fast nucleation of $Al_2O_3$. These processes may leave less oxygen for Cu NPs oxidation. In Figure 12, extinction curves for pure Cu (a) and pure Al (b) NPs colloids are calculated (dashed-dotted lines) using Mie theory and L-M[25] minimization algorithm to fit the experimental curves for Cu and Al ablated in water (full lines).

For Cu NPs, the model considers that Cu nucleates and interacts with the oxygen in the water, forming a core-shell type dimer, in agreement with the results shown in the literature.[34,37-39] For this reason a presence of $Cu_2O$ is considered in the extinction calculations. This $Cu_2O$ fraction may be included either by a mixture of dielectric functions (the well-known Maxwell-Garnett's function, MG[40,41]) or by considering a core-shell system, where $Cu_2O$ forms an outer layer coating around the Cu core. The best fit (Figure 12 (a)) is obtained considering a core-shell model, where the Cu modal core radius and the $Cu_2O$ shell thickness are found through L-M algorithm, yielding a modal core radius of 0.6 nm with a mean shell thickness of 40% core radius plus a very low contribution (~ 0.15%) of NPs with sizes larger than 50 nm with similar oxide shell thicknesses, as can be seen in the inset.

The same fitting procedure is used for Al colloids in water, where alumina is considered in the calculations since, as stated before, it is the most common and stable of all Al oxides. As in the case of Cu, core-shell structures and a mixture of dielectric functions were considered. The best result (Figure 12 (b)) was obtained for a MG type NP of alumina with 10.1 nm modal radius with 0.35% Al aggregates, also depicted in the inset.

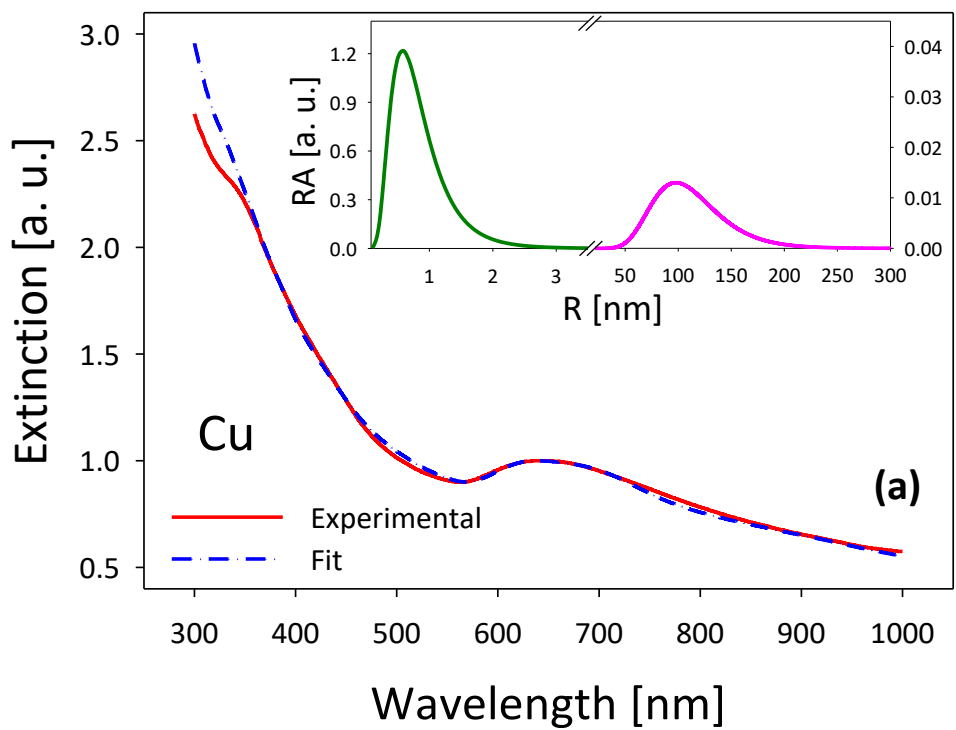

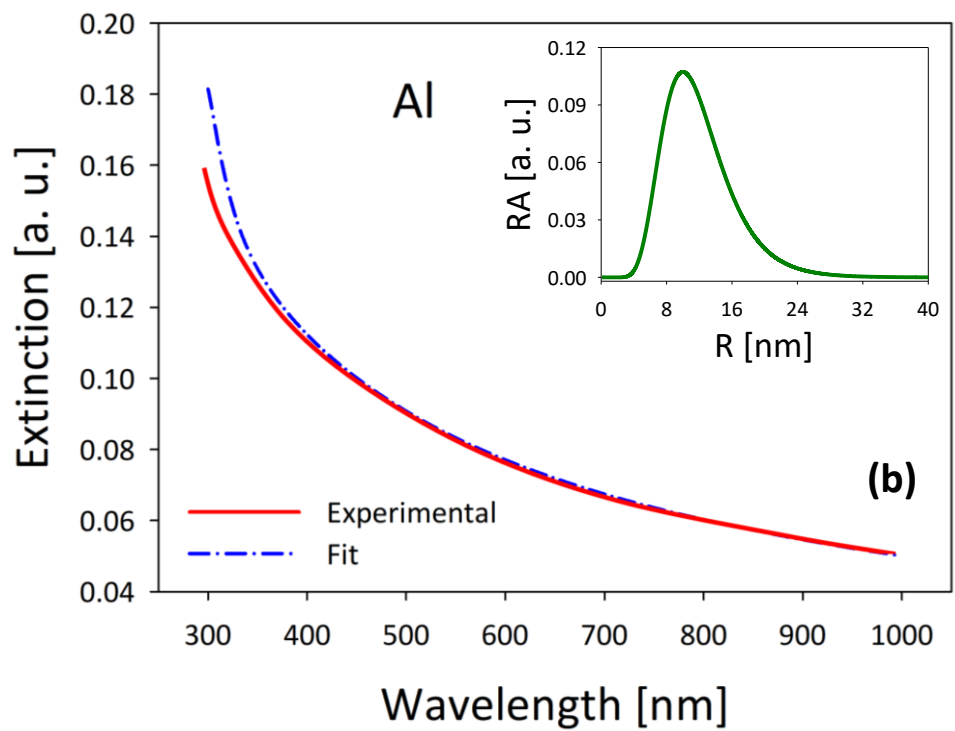

Figure 12: Extinction spectra for NPs colloids obtained from pure Cu (a) and pure Al (b) targets by FLASiS in water. Solid and dashed-dotted lines correspond to experimental data and the best fit, respectively. Insets show the Relative Abundance (RA) of the NPs within the samples for each metal. Cu colloids (a) present two populations with total (core-shell) radii of 0.84 nm and 110 nm with RA indicated in the Y left axis and Y right axis respectively. Al colloids (b) present alumina NPs with a size distribution having a modal radius of 10.1 nm.

So, concerning the case of alloy AlCu NPs synthesized by FLASiS, when the plasma formed over the target is in its high temperature phase, oxidation Al reaction proceeds faster than Cu oxidation, and Cu of mainly 1 nm radius (Figure 12 (a)) nucleates within the alumina of about 10 - 20 nm radius (Figure 12 (b)) when the plasma temperature decreases. In this way, a NP of alumina, with some inclusions of Cu, and possibly a very low proportion of Al is expected to be obtained at the end of the process.

Regarding alloy NPs extinction spectra, it is not possible to find a full wavelength range fit since the model is based on many parameters (NP size, inclusion size, filling factor and Al oxide percentage) and an absolute minimum in the L-M algorithm cannot be found in a reliable way. However, if the analysis is restrained to the plasmonic resonance region, the shape of the spectra can be approximately modelled considering an alumina NP of 10 nm radius with Cu inclusions of 1 nm radius with different $f$ values. Figure 13 shows that, for these conditions, the shape of the spectrum changes dramatically as $f$ increases. The curve for $f = 0.005$ is the best approximation to the shape of the $AlCu_{53.7\%}$ spectrum shown in Figure 11.

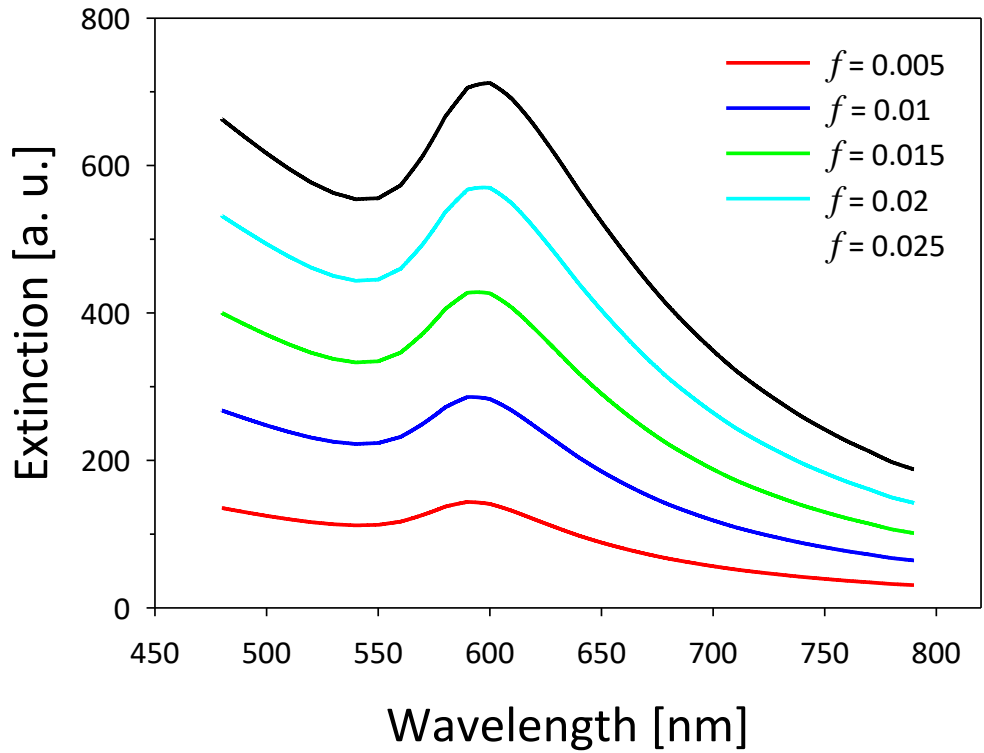

Figure 13: Extinction spectra for alumina NP of 10 nm radii with Cu inclusions of 1 nm radii for different $f$ values.

Similarly, Figure 14 shows calculated spectra for alumina NP of 20 nm radii with Cu inclusions of 1 nm radius for a lower range of $f$ values, including pure alumina ($f = 0$). In this case, the curve for $f = 0.0020$ also approximates well the shape of the $AlCu_{53.7\%}$ spectrum shown in Figure 11.

Although these calculations correspond to monodispersed NPs, according to the size distributions that best fit pure Al and Cu NPs (insets in Figure 12), they suggest that alloy NPs are mainly formed by 10 - 20 nm alumina with small filling factors of Cu NPs of approximately 1 nm radius.

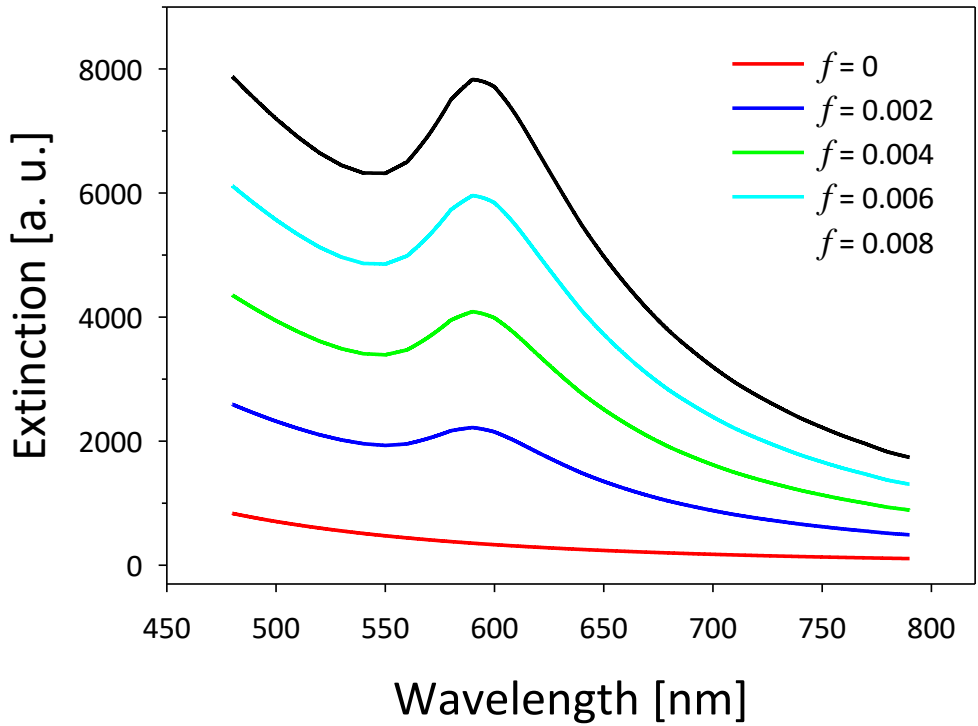

Figure 14: Extinction spectra for alumina NP of 20 nm radii with Cu inclusions of 1 nm radii for different *f* values.

Values of *f* obtained in Figures 13 and 14 are lower than those derived from EDS analysis for AlCu$_{33}$ (*f* = 0.015) in Figure 7 (c) and for AlCu$_{53.7}$ (*f* = 0.053) in Figure 8 (a) since the actual sample contains alloy NPs with sizes much smaller than the NPs selected for EDS analysis.

## 5. Conclusions

We have studied for the first time the structure of AlCu alloy NPs obtained by FLASiS of bulk target with different Cu weight percentage. SEM analysis provided a picture of the phase distribution in the bulk samples, observing Cu-rich or Al-rich island-shaped regions. Since these phases exhibit similar sizes as the laser spot, during laser ablation process the laser beam impacts over these distinct zones, sputtering out Cu or Al into the surrounding aqueous medium.

Laser ablation of these alloy targets in aqueous solution produces favourable conditions for the production of alloy NPs through a local increase of pressure and temperature in the plasma plume. The similar weight percentages of aluminium and oxygen found in these NPs suggest that Al has undergone an oxidation process due to reaction with water at this temperature. In this way, EDS analysis supports the idea of alumina NPs within which Cu nucleates forming inclusions. A model for its morphological composition is proposed to qualitatively describe the main features of the experimental spectra. The optical extinction spectra of pure Al and pure Cu NPs ablated separately in water were fitted through Mie calculations, considering that sputtered Al and Cu NPs are mainly oxidized into alumina and core-shell $Cu-Cu_2O$ respectively. When AlCu alloy targets are ablated in water, the oxidation reaction of the sputtered Al proceeds faster than that of the Cu reaction during the transient high temperature phase of the ablation plasma, leaving little oxygen for Cu oxidation. As the plasma temperature decreases, Cu nucleates within the alumina, yielding an alumina NP with some inclusions of Cu at the end of the process.

For the case of alloys, the large number of fitting parameters precludes a reliable fitting. The spectra in the region of the plasmonic band can be simulated considering alumina NPs with sizes in the range 10 - 20 nm radii with small filling factors of 1 nm copper inclusions in the range 0.001 to 0.002, which agree with the size distribution derived from pure Al and pure Cu extinction spectra.

Although it was also found that colloid NPs obtained after FLASiS does not keep the same volumetric component relation than in the original bulk target, TEM analysis and quantitative EDS spectra show that elemental composition in substructural characteristics of inclusions correspond to electron-dense zones made up of copper atoms, whose percentages depend on the initial mixture composition, meaning that when Cu quantity in bulk target becomes larger, so is its inclusion within the alumina NPs. This correlation is also observed in the increase of Cu plasmonic band in OES for colloid alloy NPs obtained from bulk samples of increasing Cu percentage.

**Acknowledgements**

This work was granted by PIP CONICET 0279, PUE CONICET 22920170100016CO (2018-2022), PIP CONICET- 11220200102846CO and PICTE- CONICET 2022-06-00322, MINCyT PME 2006-00018, 11/I237, Facultad de Ingeniería at Universidad Nacional de La Plata (UNLP), PICT ANPCyT 2018-03451. Bulk alloy disk samples were fabricated at IFIMAT (UNCPBA). Fabrication of NPs by Ultrafast Pulse Laser Ablation were carried out at CIOp (CONICET, CICBA and UNLP), La Plata. We gratefully acknowledge Prof. Dr. Alberto Caneiro from Y-TEC Argentina for the TEM and SEM images appearing in this work.

J. M. J. S., D. M. A., O. F., M. L. and L. B. S. belong to CONICET and D. C. S. belong to CICBA, Argentina